\documentclass[aps,prb,twocolumn,superscriptaddress,floatfix]{revtex4-2}
\usepackage[utf8]{inputenc}
\usepackage[english]{babel}
\usepackage{amsmath}
\usepackage{upgreek} 
\usepackage{feynmf}
\usepackage{xcolor}
\usepackage{soul}
\usepackage{graphicx}
\usepackage{amsfonts}
\usepackage{blindtext}
\usepackage{appendix}
\usepackage[colorlinks=true,unicode=true,allcolors=blue]{hyperref}
\usepackage{siunitx}
\usepackage[normalem]{ulem}
\addto\captionsenglish{}

\begin{document}

\preprint{APS/123-QED}
\title{Cavity-less Brillouin strong coupling in a solid-state continuous system}
\author{Laura Blázquez Martínez}
\affiliation{
 Max Planck Institute for the Science of Light, Staudtstr. 2, 91058, Erlangen, Germany}
 \affiliation{Department of Physics, Friedrich-Alexander Universität Erlangen-Nürnberg, Staudtstr. 7, 91058 Erlangen, Germany}
 \affiliation{Institute of Photonics, Leibniz Universität Hannover, Welfengarten 1A 30167, Hannover, Germany}
\author{Changlong Zhu}
\affiliation{
 Max Planck Institute for the Science of Light, Staudtstr. 2, 91058, Erlangen, Germany}
 \affiliation{Department of Physics, Friedrich-Alexander Universität Erlangen-Nürnberg, Staudtstr. 7, 91058 Erlangen, Germany}
\author{Birgit Stiller}
 \email[Corresponding author ]{birgit.stiller@mpl.mpg.de}
\affiliation{
 Max Planck Institute for the Science of Light, Staudtstr. 2, 91058, Erlangen, Germany}
 \affiliation{Department of Physics, Friedrich-Alexander Universität Erlangen-Nürnberg, Staudtstr. 7, 91058 Erlangen, Germany}
 \affiliation{Institute of Photonics, Leibniz Universität Hannover, Welfengarten 1A 30167, Hannover, Germany}
\date{\today}

\begin{abstract}
Strongly coupling two systems allows them to exchange coherent information before the systems decohere. This important regime in light-matter interactions has predominantly been reached in optical resonator configurations. In this work, we present the experimental realization of strong coupling between optical and acoustic fields within a continuum of modes in a cavity-less configuration after a single-pass through an optical waveguide. The underlying physical effect of anti-Stokes Brillouin-Mandelstam scattering in a highly nonlinear fiber at T~=~4~K allows us to experimentally demonstrate strong coupling in a waveguide scenario. We show the splitting of the optoacoustic spectral response and introduce a novel technique to measure the avoided crossing of hybrid optoacoustic modes via forced detuning. This demonstration opens a path towards in-line acoustic-waves-based quantum signal processing in waveguide systems.
\end{abstract}

\maketitle

\section{Introduction} 

In the strong coupling regime, energy can be exchanged coherently between two systems before the coupled system decoheres. This critical regime is in general important for any kind of coupled oscillators \cite{novotnyStrongCouplingEnergy2010, garcia-vidalManipulatingMatterStrong2021, s.dovzhenkoLightMatterInteraction2018b} and it has found particular interest in the field of light-matter interactions, described by quantum electrodynamics. 
Strong coupling has found a plethora of applications in molecular chemistry \cite{hiraiMolecularChemistryCavity2023a}, condensed matter physics \cite{khitrovaVacuumRabiSplitting2006, koppensGraphenePlasmonicsPlatform2011}, atomic physics \cite{pedrozo-penafielEntanglementOpticalAtomicclock2020}, nanoscience \cite{leeStrongCouplingPlasmonic2023, bittonQuantumDotPlasmonics2019}, photonics \cite{geStronglyCoupledSystems2021, riveraLightMatterInteractions2020}, and biomedicine \cite{kimRecentAdvancesQuantum2023}.
Especially the coherent transfer of information, which is inherent to strong coupling, has attracted a lot of interest in the field of quantum signal processing \cite{stannigelOptomechanicalQuantumInformation2012, teufelCircuitCavityElectromechanics2011b, oconnellQuantumGroundState2010a}, as quantum properties are preserved in the process. 
In the realm of optomechanics, where electromagnetic radiation couples to mechanical vibrations, light has successfully been strongly coupled to micromechanical resonators \cite{groblacherObservationStrongCoupling2009, kharelMultimodeStrongCoupling2022b, delosriossommerStrongOptomechanicalCoupling2021, verhagenQuantumcoherentCouplingMechanical2012a, tomasellaStrongOptomechanicalCoupling2025, correiaCoherentInterferometricControl2024}, Bose-Einstein condensates \cite{brenneckeCavityOptomechanicsBoseEinstein2008} and membranes \cite{thompsonStrongDispersiveCoupling2008}. Working in this regime has allowed to generate mechanical non-classical states \cite{qianQuantumSignaturesOptomechanical2012, wiseNonclassicalMechanicalStates2024a, yangMechanicalQubit2024, mayorHighPhotonphononPair2025},
measure quantum effects \cite{purdyQuantumCorrelationsRoomtemperature2017, murchObservationQuantummeasurementBackaction2008, rablPhotonBlockadeEffect2011}, influence phonon thermodynamics \cite{yangPhononHeatTransport2020, liuDynamicDissipativeCooling2013a} and improve sensing \cite{liCavityOptomechanicalSensing2021}.

The particular case of light interacting with traveling acoustic waves, described by Brillouin-Mandelstam scattering, has received interest in recent years, with applications such as sensing \cite{niklesSimpleDistributedFiber1996, geilenExtremeThermodynamicsNanolitre2023, stillerPhotonicCrystalFiber2010}, lasing \cite{otterstromSiliconBrillouinLaser2018a, hillCwGenerationMultiple1976b, zengOpticalVortexBrillouin2023} or integrated photonics \cite{marpaungIntegratedMicrowavePhotonics2019}. A wide range of optoacoustic platforms is available, both in material \cite{kittlausLargeBrillouinAmplification2016, ogusuBrillouingainCoefficientsChalcogenide2004b, ahmadUltranarrowLinewidthSingle2013, rodriguesCrossPolarizedStimulatedBrillouin2025, weiProgrammableMultifunctionalIntegrated2025} and structural \cite{chemnitzLiquidCoreOpticalFibers2023, zengStimulatedBrillouinScattering2022a, beugnotCompleteExperimentalCharacterization2007a, godetBrillouinSpectroscopyOptical2017, eggletonBrillouinIntegratedPhotonics2019,renningerBulkCrystallineOptomechanics2018a} diversity. Waveguides, with their defined direction of propagation, allow to have temporal control over the propagation of optical signals through them, which has in particular been exploited for signal processing \cite{beckerOptoacousticFieldprogrammablePerceptron2024, vidalTunableReconfigurablePhotonic2007a, liuIntegratedMicrowavePhotonic2020a, zengNonreciprocalVortexIsolator2022a,  slinkovAllopticalNonlinearActivation2025, safferBrillouinbasedStorageQPSK2025} and storage \cite{stillerCoherentlyRefreshingHypersonic2020, zhuStoredLightOptical2007, merkleinChipintegratedCoherentPhotonicphononic2017a}. 
The natural symmetry-breaking of Stokes and anti-Stokes processes has been exploited for active cooling \cite{otterstromOptomechanicalCoolingContinuous2018a, blazquezmartinezOptoacousticCoolingTraveling2024a}, without the need of working in the resolved sideband regime like in resonators \cite{teufelSidebandCoolingMicromechanical2011c, chanLaserCoolingNanomechanical2011c}. The shape of the spectrum of acoustic modes present in a waveguide is moreover crucially different from that of a resonator. In an optomechanical cavity, narrow discrete resonances are addressed, contrary to the close to continuous spectrum of modes present in a waveguide. 

Reaching the strong coupling regime via Brillouin-Mandelstam scattering has been theoretically studied \cite{zhangQuantumCoherentControl2023b, huyStrongCouplingPhonons2016b} and pioneering experimental work has been shown in resonators \cite{enzianObservationBrillouinOptomechanical2019b, wangTamingBrillouinOptomechanics2024} and superfluid helium \cite{heStrongOpticalCoupling2020b}. All of the previous optomechanical and optoacoustic approaches have in common that the experiments were performed in cavity configurations, where the driving optical field resonates in an optical cavity. Light strongly couples to the medium, either to distinct mechanical resonances or traveling waves. So far, strong coupling in the simpler but substantially more challenging experimental configuration of an optical waveguide \cite{rakichQuantumTheoryContinuum2018b} has not been demonstrated.

\begin{figure*}
	\centering
	\includegraphics[width=0.95\textwidth]{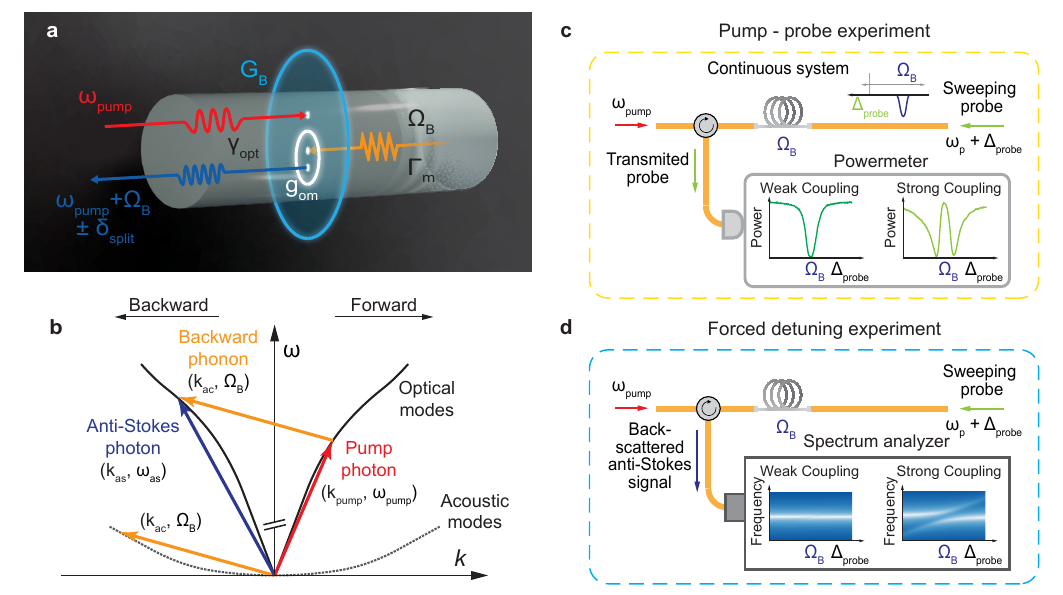} 
	\caption{\textbf{Cavity-less strong coupling and how to measure it.} (a)~Sketch of an anti-Stokes Brillouin-Mandelstam scattering process in the strong coupling regime. (b)~Dispersion diagram for the different modes involved in an anti-Stokes Brillouin-Mandelstam scattering process. Adapted from \cite{wolffBrillouinScatteringTheory2021c, enzianObservationBrillouinOptomechanical2019b}. (c)~Schematic of a pump-probe experiment. (d)~Schematic of a forced detuning experiment.}
	\label{fig:Fig1_Concept}
\end{figure*}

In this work, we report the experimental realization of cavity-less optoacoustic strong coupling between optical photons and traveling acoustic phonons via Brillouin-Mandelstam scattering. We experimentally demonstrate that optical radiation - after a single pass through a waveguide - strongly couples to acoustic phonons within a continuum of modes. Contrary to other cases of light-matter interaction, we reach the strong coupling regime both in the absence of an optical cavity and without coupling to discrete resonances of a system, such as in a mechanical resonator. In previously reported experiments without an optical cavity, light is either coupled to discrete vibrational modes of molecules \cite{thomasCavityFreeUltrastrongLightMatter2021}, via excitons to discrete mechanical resonances \cite{okamotoCavitylessOnchipOptomechanics2015b, changCavityOptomechanicsCavityless2022} or to semiconductors \cite{muckeSignaturesCarrierWaveRabi2001}. We were able to access the strong coupling regime due to our experimental configuration balancing Brillouin gain, optical power handling and optical and mechanical dissipation rates. In the experimental results, we first demonstrate the splitting of the optoacoustic spectral response. We moreover introduce a new measurement technique to observe the avoided crossing of hybrid optoacoustic modes via forced detuning. The experimental results are confirmed by our analytical theory on dynamic continuum quantum optoacoustics.

\section*{Experimental methods}

Using Brillouin-Mandelstam scattering, back-scattered anti-Stokes photons are strongly coupled with counter-propagating thermal acoustic phonons (Fig.~\ref{fig:Fig1_Concept}~(a)). The platform chosen for this experiment is a highly nonlinear fiber (HNLF) in a 4~K cryostat. The HNLF has the same core-cladding structure as a single mode fiber (SMF), but with a higher doping of germanium in the core, where the optical loss is $\gamma_{{\rm dB}}$~=~0.001~dB/m and the material refractive index $n$~=~1.48. The material nonlinearity is increased, but the sample can be spliced to SMF with low loss, resulting in a optical dissipation rate of $\gamma_{{\rm opt}}$~=~24~±~2~MHz. More information about the sample can be found in supplementary materials. Performing the experiments at cryogenic temperature presents the advantage of an increased Brillouin gain in the fiber, from $G_{\rm B}$(293~K)~=~1.60~±~0.06~${\rm W^{-1}m^{-1}}$ to $G_{\rm B}$(4~K)~=~6~±~1~${\rm W^{-1}m^{-1}}$. Moreover, the mechanical dissipation rate to the thermal bath $\Gamma_{\rm m}$ decreases at cryogenic temperatures \cite{cryer-jenkinsBrillouinMandelstamScattering2025b}, from $\Gamma_{\rm m}$(293~K)~=~46.3~±~0.2~MHz to $\Gamma_{\rm m}$(4~K)~=~9.91~±~0.07~MHz. These material changes at cryogenic temperature, combined with the low optical dissipation rate of the sample and its capability to handle elevated optical pump powers ($<~2$~W) to access high coupling values allow us to enter the strong coupling regime.

\begin{figure} 
	\centering
	\includegraphics[width=0.45\textwidth]{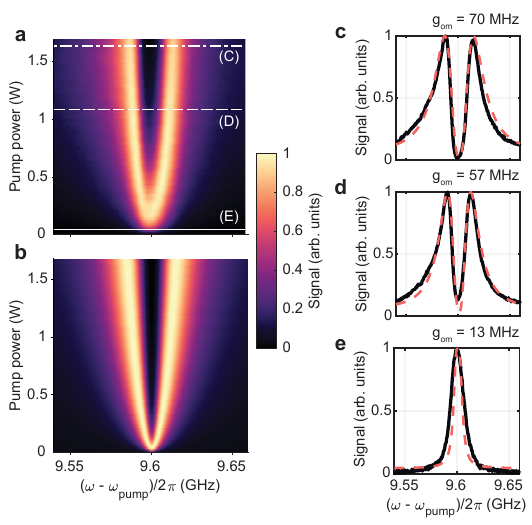} 
	\caption{\textbf{Strongly coupled Brillouin response of a $T$~=~4~K fiber.} (a)~Anti-Stokes Brillouin resonance as function of pump power. (b)~Simulation of cavity-less optoacoustic strong coupling \cite{zhangQuantumCoherentControl2023b}, showing good agreement with the experiment. (c) and (d)~Optical anti-Stokes resonance in the strong coupling regime. (e)~Optical anti-Stokes resonance in the weak coupling regime.  Experimental data shown as a dark continuous line, theoretical data shown as a light dashed line.}
	\label{fig:Fig2_Seeded} 
\end{figure}

An important characteristic of backward Brillouin-Mandelstam scattering in waveguides is the automatic phase-matching between the optical pump wave and a subset of the continuum of acoustic modes in the medium (Fig.~\ref{fig:Fig1_Concept}~(b)). This subset is given by the Lorentzian Brillouin gain spectrum \cite{wolffBrillouinScatteringTheory2021c}
\begin{equation}
     g_{\rm B}(\omega)=G_{\rm B}\frac{(\Gamma_{\rm m}/2)^2}{(\omega-\Omega_{\rm B})^2+(\Gamma_{\rm m}/2)^2},
\end{equation}
where the phonons at frequency $\Omega_{\rm B}$ are perfectly phase-matched and the linewidth $\Gamma_{\rm m}$ defines the limit of quasi-phase-matched phonons addressed in the interaction. As the gain spectrum is a subset of the acoustic continuum, the system is always in resonance, as long as the pump wavelength is inside the optical transparency window of the material. Shifting $\lambda_{\rm pump}$ results only in a shift of the measured acoustic resonance frequency $\Omega_{\rm B}~=~4\pi nv_{\rm ac}/\lambda_{\rm pump}$ \cite{wolffBrillouinScatteringTheory2021c} (where $v_{\rm ac}$ is the acoustic velocity), as the resonant acoustic modes are different. The coupling of the interaction is defined as  $g_{\rm om}\approx\sqrt{G_{\rm B}\Gamma_{\rm m}P_{\rm pump}c/4n}$, where $c$ is the speed of light in vacuum and $P_{\rm pump}$ the pump power \cite{kharelNoiseDynamicsForward2016a, zhangQuantumCoherentControl2023b}. In the case of resonators, the intracavity power increases as the drive becomes resonant with the cavity. Following a mechanical eigenfrequency of the optomechanical system, a splitting is observed when entering the strong coupling regime. At this point, called avoided crossing, the approximation of weakly coupled optical and mechanical fields breaks down and the resonance changes to those of the hybrid optomechanical system. Thus, the single Lorenztian resonance splits in two Lorentzians peaks, representing the symmetric and anti-symmetric optomechanical modes. In the case of waveguides however, a detuning from resonance is not straightforward. 

Standard detection techniques in Brillouin-Mandelstam scattering do not allow to observe an avoided crossing, as the back-scattering spectrum is usually measured in resonance. The first technique used in this paper, a pump-probe experiment \cite{wolffBrillouinScatteringTheory2021c}, is shown in Fig.~\ref{fig:Fig1_Concept}~(c). A Brillouin resonance is excited by the pump and a counter-propagating low power probe sweeps over it ($ \omega_{\rm probe}=\omega_{\rm pump}+\Delta_{\rm probe}$). As the anti-Stokes process is a phononic loss channel, the probe is depleted. The transmitted probe power depends of the detuning $\Delta_{\rm probe}$, therefore the frequency information to obtain a spectrum comes from the probe sweep. Note here that, even if the probe can be off-resonance, the resulting spectrum from this measurement will be in resonance, as each probe detuning corresponds to a single transmitted power. An alternative technique is to perform a thermal-noise-initiated measurement. Here, solely the pump is launched into the sample, and the back-reflected scattered signal detected by an spectrometer. As neither of these techniques are apt for detuning measurements, we introduce a novel experimental technique, shown in Fig.~\ref{fig:Fig1_Concept}~(d). It combines the frequency scanning of a pump-probe experiment with spectral detection to measure off-resonance spectra. Via this forced detuning, the avoided crossing between two strongly coupled optoacoustic modes can be observed in the absence of a cavity. 

\section*{In-resonance measurement of strong coupling}

As stated, standard techniques measure in waveguides the resonant backward scattering spectrum. To reproduce the effect of increasing intracavity power as the optical drive approaches resonance, a pump power sweep is performed. The experiment realized is a pump-probe measurement, where a low power anti-Stokes probe of frequency $\omega_{{\rm probe}}~=~\omega_{{\rm p}}+\Omega_{\rm B}+\Delta_{{\rm probe}}$ is used. The results are equivalent to those from a thermal-noise-initiated experiment (see supplementary materials). Performing a pump-probe measurement, combined with lock-in amplifier detection, increases significantly the signal-to-noise ratio, while minimizing the signal contribution from sections of the sample not thermalized to 4~K (see supplementary materials). The results are shown in Fig.~\ref{fig:Fig2_Seeded}~(a). 

For low input powers, the system is weakly coupled ($g_{{\rm om}}$~=~13~±~2~MHz) and the anti-Stokes resonance has the shape of a single Lorenzian centered at $\Omega_{\rm B}/2\pi$~=~9.5997~±~0.0001~GHz (Fig.~\ref{fig:Fig2_Seeded}~(e)). As the pump power increases ($g_{{\rm om}}$~=~57~±~2~MHz), the system enters the strong coupling regime and the resonance splits (Fig.~\ref{fig:Fig2_Seeded}~(d)). Continuing to increase the pump power pushes the system deeper into the strong coupling regime ($g_{{\rm om}}$~=~70~±~2~MHz, Fig.~\ref{fig:Fig2_Seeded}~(c)), resulting in a maximum measured split of $\delta_{{\rm split}}$~=~26.7~MHz for $g_{{\rm om}}$~=~71~±~2~MHz. The shape of these two peaks is asymmetric, with the outer slopes of each peak following the shape of a Lorentzian slope, while the inner slopes are sharper. In order to validate our results, we compare with our theoretical model, which shows good agreement with the experiment (Fig.~\ref{fig:Fig2_Seeded}~(b)).

\begin{figure*} 
	\centering
	\includegraphics[width=0.9\textwidth]{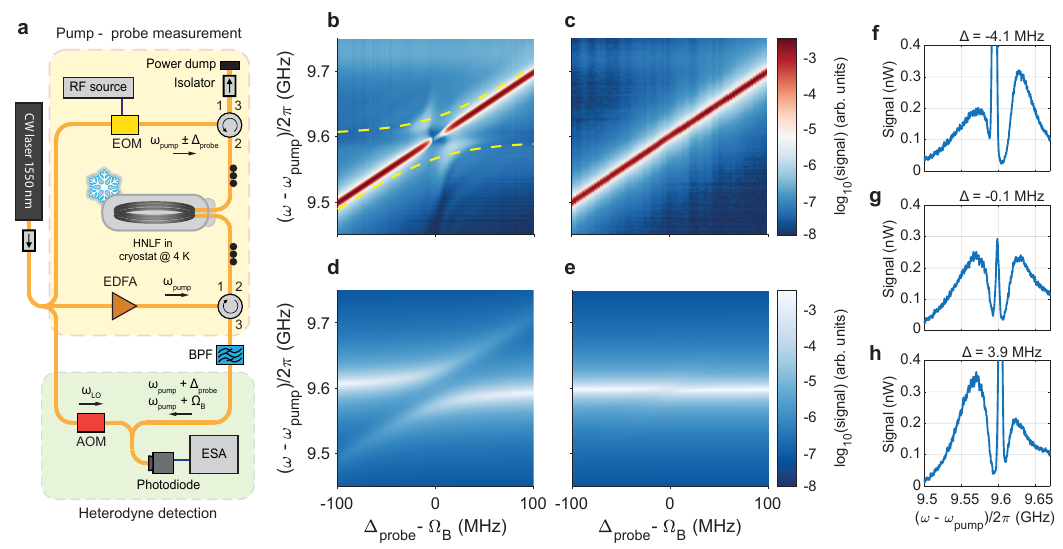}
	\caption{\textbf{Forced detuning experiment and results.} (a)~Diagram of the setup to perform a forced detuning measurement. CW: continuous wave; EDFA: Erbium-doped fiber amplifier; BPF: band-pass filter; E(A)OM: electro(acusto)optical modulator; ESA: electrical spectrum analyzer. (b)~Experimental realization of a forced detuning measurement using a low power probe (P$_{{\rm probe}}$~=~0.6~µW) in the strong coupling regime ($g_{{\rm om}}$~=~64~±~2~MHz, $P_{{\rm pump}}$~=~1.35~±~0.1~W, $T~\approx$~4~K). Yellow dashed line represents the optoacoustic eigenvalues. (c) Forced detuning measurement in the weak coupling regime ($g_{{\rm om}}$~=~12~±~2~MHz, $P_{{\rm pump}}$~=~0.05~±~0.1~W, $T~\approx$~4~K). (d)~Theoretical simulation of an anti-crossing measurement in a cavity-less optoacoustic system, showing good agreement with the experiment. (e)~Same as (d), for the weak coupling case. (f)~Spectrum in the strong coupling regime with a red detuned off-resonance probe, (g)~with a resonant probe, (h)~with a blue detuned off-resonance probe.}
	\label{fig:Fig3_HeteroSeeded} 
\end{figure*}

In a cavity-less system, at position $z$ along the 1-D propagation direction, the envelope bosonic operators of the optical anti-Stokes and acoustic waves can be constructed as 
\begin{equation}
	a_{\rm aS} = 1/\sqrt{2\pi}\int dk a(k,t) e^{-ikz},
	\label{eq:Envelope_aS} 
\end{equation}
\begin{equation}
	b_{\rm ac} = 1/\sqrt{2\pi}\int dk b(k,t) e^{-ikz},
	\label{eq:Envelope_aS} 
\end{equation}
respectively, where $a(k,t)$ and $b(k,t)$ denote the annihilation operators for the $k$-th anti-Stokes photon mode and acoustic phonon mode with wave number $k$. In the undepleted pump regime, the interaction Hamiltonian can be linearized to a beam-splitter interaction 
\begin{equation}
	H_{\rm int}=\hbar g_{\rm om}\left ( a(k)^{\dagger}b(k) + b(k)^{\dagger}a(k) \right ),
	\label{eq:Hamiltonian} 
\end{equation}
with an effective pump-enhanced coupling strength $g_{\rm om}$, which generates state swapping between the two photon and phonon modes. Normal-mode splitting, one of the footprints of strong coupling, becomes observable when $2 g_{\rm om}>(\gamma_{\rm opt}+\Gamma_{\rm m})/2=17$~MHz, meaning that the coupling strength exceeds the average linewidth. Under this condition, two peaks are distinguishable in both optical and acoustic displacement spectra, as the normal modes shift to describe the hybrid optoacoustic system. This matches the experimental observations in Fig.~\ref{fig:Fig2_Seeded}~(c) and (d) (theory in dashed lines), where the displacement spectrum of the anti-Stokes wave versus the wave number $k$ shows a split. Here, $\omega_{\rm probe}-\omega_{\rm pump}=k\upsilon_{\rm opt}$ corresponds to the wave number-induced frequency shift of the anti-Stokes mode, where $\upsilon_{\rm opt}$ denotes the group velocity of the anti-Stokes wave. This resonance splitting is the waveguide equivalent of an avoided crossing in resonators.

\section*{Forced detuning measurement}

In order to measure off-resonance spectra to observe an avoided crossing and confirm our results from the previous section, a forced detuning experiment is performed. A diagram of the setup used is shown in Fig.~\ref{fig:Fig3_HeteroSeeded}~(a). The emission of a $\lambda_{\rm pump}$~=~1550~nm laser is divided in three paths, local oscillator (LO), pump and probe. The probe is modulated sinusoidally via an electro-optical modulator (EOM), generating the sideband $\omega_{\rm pump}+\Delta_{\rm probe}$, which is swept around the Brillouin resonance $\Omega_{\rm B}$. The light in the pump arm is amplified via an Erbium-doped fiber amplifier (EDFA), to bring the system into the strong coupling regime. Both high power pump and sweeping probe are sent to the cryogenic sample, resulting in a signal coming out of port 3 of the pump circulator containing the spontaneous backscattered Brillouin response and transmitted probe. This signal is filtered around the anti-Stokes response and detected via heterodyne detection on a photodiode connected to an electrical spectrum analyzer (ESA).

An avoided crossing is successfully observed in Fig.~\ref{fig:Fig3_HeteroSeeded}~(b). A low power probe (red diagonal) is swept over the strongly coupled resonance, spontaneously generated by the pump. The probe is depleted to a great degree, confirming that this is an anti-Stokes resonance and coming from a  high gain Brillouin interaction. As different off-resonance modes are probed, energy is transferred from the upper ($\Omega_{\rm B}+\delta_{\rm split}$) to the lower frequency ($\Omega_{\rm B}-\delta_{\rm split}$) optoacoustic hybrid mode. A dashed yellow line representing the two hybrid eigenvectors is added as a visual aid, as the presence of the probe and the resonances stemming from the warm part of the fiber create a high background. This dashed line is obtained from the theoretical description presented in this work (see supplementary materials), which allows to reproduce the full spectrum of the avoided crossing, as shown in Fig.~\ref{fig:Fig3_HeteroSeeded}~(d), in good agreement with the experiment. The probe shows as a diagonal in Fig.~\ref{fig:Fig3_HeteroSeeded}~(b). No probe is present in the theoretical spectrum, as it is only needed in the experiment and therefore, the diagonal does not appear in the simulation in Fig.~\ref{fig:Fig3_HeteroSeeded}~(d) and (e).

Different results are obtained when performing this measurement in the weak coupling regime, shown in Fig.~\ref{fig:Fig3_HeteroSeeded}~(c). As the interaction at $g_{{\rm om}}$~=~12~±~2~MHz has a lower gain, the probe is barely depleted. No splitting is observed and thus no avoided crossing can be seen. This is supported by the theoretical spectrum (Fig.~\ref{fig:Fig3_HeteroSeeded}~(e)), where only the resonance at $\Omega_{\rm B}$ is present, independent of $ \Delta_{\rm probe}$. As the system is weakly coupled, the energy stays in the detected higher energy anti-Stokes optical mode. 

\begin{figure} 
	\centering
	\includegraphics[width=0.45\textwidth]{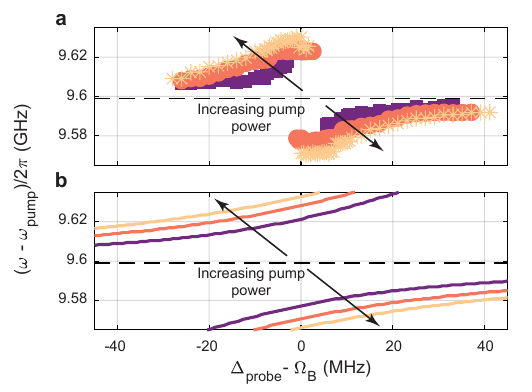} 
	\caption{\textbf{Hybrid optoacoustic mode separation versus input power.} (a)~Maxima of the measured strongly coupled optoacoustic modes as a function of pump power. Dark squares:~$g_{{\rm om}}$~=~48~±~2~MHz, medium circles:~$g_{{\rm om}}$~=~61~±~2~MHz, light stars:~$g_{{\rm om}}$~=~75~±~2~MHz. (b)~Maxima of the measured strongly coupled optoacoustic modes as a function of pump power according to the theory. Colors represent same input powers as in (a). Dashed line indicates $\Omega_{\rm B}$.}
	\label{fig:Fig4_Splitting} 
\end{figure}

Having a closer look around the resonance in the strong coupling regime allows to study the energy distribution between hybrid modes. For a red detuned probe $\Delta_{{\rm probe}}-\Omega_{\rm B}$~=~-4.1~MHz, more energy is present in the higher frequency mode (Fig.~\ref{fig:Fig3_HeteroSeeded}~(f)). When the probe is resonant $\Delta_{{\rm probe}}-\Omega_{\rm B}$~=~-0.1~MHz, the amount of energy between the two modes is the same, resulting in symmetric peak heights (Fig.~\ref{fig:Fig3_HeteroSeeded}~(g)). This matches with what is observed in the pump-probe experiment, where the measurements are performed with the system in resonance. For a blue detuned probe $\Delta_{{\rm probe}}-\Omega_{\rm B}$~=~3.9~MHz, the energy is transferred to the lower frequency mode (Fig.~\ref{fig:Fig3_HeteroSeeded}~(h)). 
This exchange of energy follows the expected behavior of strongly coupled resonances as a function of detuning. The effect of pump power is shown in Fig.~\ref{fig:Fig4_Splitting}~(a). As the pump power increases, so does the coupling strength between modes, pushing the system further into the strong coupling regime. The measured gap between modes at the avoided crossing therefore broadens, which is matched by the theoretical model (Fig.~\ref{fig:Fig4_Splitting}~(b)). Increasing the temperature of the fiber from $T$~=~4~K has the opposite effect. For the same pump power, $\Gamma_{\rm m}$ is higher and $G_{\rm B}$ lower, resulting in a higher threshold to access the strong coupling regime (see the supplementary material for more information).

\section*{Conclusions and outlook}

In this work we present what is to our knowledge the first realization of optoacoustic strong coupling in a waveguide.
We show that the experimental complexity of working with optomechanical cavities can be circumvented, presenting a system that does not need extreme low temperatures or precise control of detuning.
Moreover, our system by nature is always in resonance across the transmission window of silica glass. 
This broadband operation, in contrast to the narrow optical resonances of a high-Q cavity, is one of the characteristics that makes Brillouin-Mandelstam scattering a powerful tool for signal processing.
\\
The normal mode splitting and anti-crossing observed in our experiment stem from strong coupling between groups of optical photons and acoustic phonons at $4$\,K, where the coupling strength $2 g_{\rm om}$ exceeds the linewidths of photons and phonons.
With a high acoustic frequency of $9.6~$GHz, the thermal phonon occupation can be cooled to around 10 phonons at such a low temperature. Under these conditions, the coupled system has the potential to enter into the quantum coherent coupling regime~\cite{verhagenQuantumcoherentCouplingMechanical2012a} when increasing the pump power to around $5~$W. In this case, the coupling strength would exceed all thermal decoherence rates. 
This means that an optical waveguide with a higher Brillouin gain or lower acoustic loss, such as chalcogenide fibers~\cite{blazquezmartinezOptoacousticCoolingTraveling2024a}, is capable of realizing nonclassical states transfer between photons and phonons at a moderate cryogenic temperature of several Kelvins.
Instead of establishing strong coupling between photons and phonons with single modes~\cite{groblacherObservationStrongCoupling2009} or multiple modes~\cite{kharelMultimodeStrongCoupling2022b}, here we achieve strong coupling over a wide bandwidth of anti-Stokes photons and acoustic phonons with continuum modes, which leads to a broad range of applications, such as quantum information processing~\cite{stannigelOptomechanicalQuantumInformation2012}, entanglement distribution ~\cite{zivariOnchipDistributionQuantum2022}, quantum computing~\cite{deleonMaterialsChallengesOpportunities2021}, and precision measurements~\cite{degenQuantumSensing2017}. 
Our experiment results represent a critical step for the coherent control of photon-phonon interaction in continuum optoacoustic systems and can establish waveguide platforms as important candidates for novel quantum technologies, including nonclassic states preparation~\cite{yangMechanicalQubit2024}, quantum transduction~\cite{laukPerspectivesQuantumTransduction2020}, quantum memory~\cite{wallucksQuantumMemoryTelecom2020}), and quantum sensors~\cite{degenQuantumSensing2017}.
With the potential to be able to work in a regime of coherent information transfer, we specifically open up the path to achieve single-pass, in-line continuous variable quantum signal processing and storage.

\begin{acknowledgments}
We thank G. Leuchs, M. Sondermann, F. Marquardt, and G. Jara-Schulz for useful discussions and insightful comments. We thank A. Geilen, Z. O. Saffer, L. Fischer, P. Wiedemann, G. Slinkov, J. H. Marines Cabello, R. Jadhav and X. Zeng for their help at different stages of the experiment.

L.B.M., C.Z. and B.S. discussed the general ideas and conceived the experiment. L.B.M. built the experimental setups, conceptualized the forced detuning measurement scheme, performed measurements and analyzed the data. C.Z. derived the theoretical models and performed simulations of the experiments. B.S. supervised the project and conceived the project. All authors discussed the results. L.B.M., C.Z. and B.S. wrote the manuscript together.

The authors acknowledge funding from the Max Planck Research Groups Scheme from the Max Planck Society and the Max Planck School of Photonics.

Correspondence and requests for materials should be addressed to Birgit Stiller. The datasets generated during and/or analysed during the current study are available from the corresponding author on reasonable request.

\end{acknowledgments}

\clearpage

\newpage

\appendix
\onecolumngrid

\section*{Supplementary Materials}

\subsection*{Theoretical model for the optoacoustic continuum system}

Backward Brillouin anti-Stokes scattering in a typical optical waveguide is an optoacoustic interaction between an optical pump wave, a scattered optical anti-Stokes wave, and an acoustic wave. The dynamics of this optoacoustic interaction can be given by
\begin{eqnarray}\label{Dynamics of Brillouin anti-Stokes scattering}
	\frac{\partial a_\text{p}}{\partial t} + \upsilon_\text{opt}\frac{\partial a_\text{p}}{\partial z} &=&
	-\frac{\gamma_{\rm opt}}{2} a_\text{p} - i g_0 a_\text{aS}^{\dagger} b_\text{ac} + \sqrt{\gamma_{\rm opt}}\xi_\text{p},\nonumber\\
	\frac{\partial a_\text{aS}}{\partial t} - \upsilon_\text{opt}\frac{\partial a_\text{aS}}{\partial z} &=&
	-\frac{\gamma_{\rm opt}}{2} a_\text{aS} - i g_0 a_\text{p} b_{\text{ac}} + \sqrt{\gamma_{\rm opt}}\xi_\text{aS}, \nonumber\\
	\frac{\partial b_\text{ac}}{\partial t} - \upsilon_\text{ac}\frac{\partial b_\text{ac}}{\partial z} &=&
	-\frac{\Gamma_{\rm m}}{2}b_\text{ac} - i g_0 a_\text{p}^{\dagger} a_\text{aS} + \sqrt{\Gamma_{\rm m}}\xi_\text{ac},
\end{eqnarray}
where $a_{\rm p}(z,t)$, $a_{\rm aS}(z,t)$, and $b_{\rm ac}(z,t)$ correspond to the envelope bosonic operators of
the pump, anti-Stokes, and acoustic waves. $\gamma_{\rm opt}$ ($\Gamma_{\rm m}$) and $\upsilon_{\rm opt}$ ($\upsilon_{\rm ac}$) 
denote the linewidth and group velocity for the optical (acoustic) wave, respectively. $g_0$ is the vacuum coupling strength.
$\xi_\text{p}$, $\xi_\text{aS}$, and $\xi_\text{ac}$ are quantum Langevin noises corresponding to these three waves,
which obey the following statistical properties
\begin{eqnarray}\label{Noise correlation in spatial space}
	\langle \xi_\text{\rm p}(t,z)\rangle &=& \langle \xi_\text{\rm aS}(t,z) = \langle \xi_\text{\rm ac}(t,z)= 0, \nonumber\\
	\langle \xi_\text{\rm p}^{\dagger} (t_1,z_1)\xi_\text{\rm p}(t_2,z_2)\rangle &=&
	\langle \xi_\text{\rm aS}^{\dagger}(t_1,z_1)\xi_\text{\rm aS}(t_2,z_2)\rangle = 0, \nonumber\\
	\langle \xi_\text{\rm ac}^{\dagger}(t_1,z_1)\xi_\text{\rm ac}(t_2,z_2)\rangle &=& n_\text{th}\delta(t_1-t_2)\delta(z_1-z_2),
\end{eqnarray}
where  $n_\text{th}=1/(e^{\hbar\Omega_\text{ac}/k_\text{B} T}-1)$ denotes the thermal phonon occupation
of the acoustic wave at temperature $T$. 
In the undepleted pump regime where the reduction of pump power can be ignored, the three-wave interaction can be
reduced a linearized beamsplitter interaction between anti-Stokes photons and acoustic phonons with an effective pump-enhanced 
coupling strength $g_{\rm om}=g_0\sqrt{\langle a_{\rm p}^{\dagger} a_{\rm p} \rangle }$. Therefore, the dynamics of
the linearized optoacoustic interaction can be expressed as follows
\begin{eqnarray}\label{Dynamics of linearized Brillouin anti-Stokes scattering}
	\frac{\partial a_{\rm aS}}{\partial t} - \upsilon_{\rm opt} \frac{\partial a_{\rm aS}}{\partial z} &=&
	- \frac{\gamma_{\rm opt}}{2} a_{\rm aS} - i g_{\rm om} b_{\rm ac} + \sqrt{\gamma_{\rm opt}}\xi_{\rm aS}, \nonumber\\
	\frac{\partial b_{\rm ac}}{\partial t} - \upsilon_{\rm ac} \frac{\partial b_{\rm ac}}{\partial z} &=&
	- \frac{\Gamma_{\rm m}}{2} b_{\rm ac} - i g_{\rm om} a_{\rm aS} + \sqrt{\Gamma_{\rm m}} \xi_{\rm ac}.
\end{eqnarray}
Without the cavity structure, the optical anti-Stokes and acoustic waves contains continuously accessible groups
of optical photons and acoustic phonons, respectively, i.e., envelope bosonic operators $a_{\rm aS}(z,t)$ and
$b_{\rm ac}(z,t)$ at position $z$ along the 1D waveguide can be constructed as
\begin{eqnarray}
a_{\rm aS}(z,t) = \frac{1}{2\pi}\int dk~a(k,t) e^{-ikz}, \quad 
b_{\rm ac}(z,t) = \frac{1}{2\pi}\int dk~b(k,t) e^{-ikz},
\end{eqnarray}
where $a(k,t)$ and $b(k,t)$ are annihilation operators for the $k$-th Stoke photon mode and acoustic phonon mode
with wave number $k$. Moving into momentum space, Eq.~(\ref{Dynamics of linearized Brillouin anti-Stokes scattering})
can be rewritten as
\begin{eqnarray}\label{Dynamics of linearized Brillouin anti-Stokes scattering in K space}
	\frac{d a(k,t)}{dt} &=& \left( -\frac{\gamma_{\rm opt}}{2}+i\Delta_{\rm as} \right)a - i g_{\rm om} b + \sqrt{\gamma_{\rm opt}}\xi_{a}, \nonumber\\
	\frac{d b(k,t)}{dt} &=& \left( -\frac{\Gamma_{\rm m}}{2}+i\Delta_{\rm ac} \right)b - i g_{\rm om} a + \sqrt{\Gamma_{\rm m}}\xi_{b},
\end{eqnarray}
where $k$ is the wave number in the momentum space. We can see that Eq.~(\ref{Dynamics of linearized Brillouin anti-Stokes scattering in K space})
describes a beamsplitter interaction between the $k$-th photon and phonon modes with interaction Hamiltonian 
$H_{\rm int}=\hbar g_{\rm om}(a^{\dagger}(k,t)b(k,t)+b^{\dagger}(k,t)a(k,t))$. Moving into the frequency domain
via a Fourier transformation, Eq.~(\ref{Dynamics of linearized Brillouin anti-Stokes scattering in K space}) can be
re-expressed as 
\begin{eqnarray}
	-i\omega \tilde{a}(k,\omega) &=& \left( -\frac{\gamma_{\rm opt}}{2} + i\Delta_{\rm as} \right) \tilde{a}(k,\omega) - i g_{\rm om}\tilde{b}(k,\omega) + \sqrt{\gamma_{\rm opt}}\tilde{\xi}_{a}(k,\omega), \nonumber\\
	-i\omega \tilde{b}(k,\omega) &=& \left( -\frac{\Gamma_{\rm m}}{2} + i\Delta_{\rm ac} \right) \tilde{b}(k,\omega) - i g_{\rm om}\tilde{a}(k,\omega) + \sqrt{\Gamma_{\rm m}}\tilde{\xi}_{b}(k,\omega),
\end{eqnarray}
where $\tilde{a}$, $\tilde{b}$, $\tilde{\xi}_{a}$, and $\tilde{\xi}_{b}$ correspond to the Fourier transform of
$a$, $b$, $\xi_{a}$, and $\xi_{b}$. The solutions of above equations can be given by
\begin{eqnarray}\label{Solutions of a and b in frequency domain}
	\tilde{a}(k,\omega) &=& \frac{ \left( \frac{\Gamma_{\rm m}}{2} - i(\omega+\Delta_{ac})  \right)\sqrt{\gamma_{\rm opt}}\tilde{\xi}_{a}(\omega) 
		- i g_{\rm om}\sqrt{\Gamma_{\rm m}}\tilde{\xi}_{b}(\omega) }
	{ g_{\rm om}^2 + \left( \frac{\gamma_{\rm opt}}{2} - i(\omega+\Delta_{\rm as}) \right) \left( \frac{\Gamma_{\rm m}}{2} - i(\omega+\Delta_{\rm ac}) \right) }, \nonumber\\
	\tilde{b}(k,\omega) &=& \frac{ -i g_{\rm om}\sqrt{\gamma_{\rm opt}}\tilde{\xi}_{a}(\omega) + \left( \frac{\gamma_{\rm opt}}{2} 
		- i(\omega+\Delta_{\rm as}) \right)\sqrt{\Gamma_{\rm m}}\tilde{\xi}_{b}(\omega) }{ g_{\rm om}^2 + \left( \frac{\gamma_{\rm opt}}{2} - i(\omega+\Delta_{\rm as}) \right) \left( \frac{\Gamma_{\rm m}}{2} - i(\omega+\Delta_{\rm ac}) \right) }.
\end{eqnarray}
We define the quadrature operators and the corresponding Fourier transform of anti-Stokes and acoustic modes as follows
\begin{eqnarray}
	X_{a}(k,t) &=& \frac{ a^{\dagger}(k,t) + a(k,t) }{\sqrt{2}}, 
	\quad Y_{a}(k,t) = \frac{ a(k,t) - a^{\dagger}(k,t) }{\sqrt{2}i}, \nonumber\\
	\tilde{X}_{b}(k,t) &=& \frac{ \tilde{b}^{\dagger}(k,t) + \tilde{b}(k,t) }{\sqrt{2}}, \quad 	
	\tilde{Y}_{b}(k,t) = \frac{\tilde{b}^{\dagger}(k,t) - \tilde{b}(k,t)  }{\sqrt{2}i}.
\end{eqnarray}
Thus we have
\begin{eqnarray}\label{Displacement spectrum of anti-Stokes 1}
	\langle \tilde{X}_{a}^{\dagger}(\omega_1) \tilde{X}_{a}(\omega_2) \rangle
	= 2\pi S_{XX}^{\rm as}(\omega_1) \delta(\omega_1-\omega_2),
\end{eqnarray}
where $S_{XX}^{\rm as}(k,\omega)$ denotes the displacement spectrum for the $k$-th anti-Stokes mode.
Based on Eq.~(\ref{Noise correlation in spatial space}), the Langevin noises of $\tilde{\xi}_a$ and
$\tilde{\xi}_b$ obey correlation relationships as follows
\begin{eqnarray}\label{Noise correlation in momentum space}
	\langle \tilde{\xi}^{\dagger}_a (k,\omega_1) \tilde{\xi}_a (k,\omega_2) \rangle &=& 0, \nonumber\\
	\langle \tilde{\xi}_a (k,\omega_1) \tilde{\xi}_{a}^{\dagger} (k,\omega_2) \rangle &=& 2\pi \delta(\omega_1-\omega_2), \nonumber\\
	\langle \tilde{\xi}^{\dagger}_{b} (k,\omega_1) \tilde{\xi}_{b} (k,\omega_2) \rangle &=& 2\pi n_{\rm th} \delta(\omega_1-\omega_2), \nonumber\\
	\langle \tilde{\xi}_b (k,\omega_1) \tilde{\xi}^{\dagger}_b (k,\omega_2) \rangle &=& 2\pi(n_{\rm th}+1)\delta(\omega_1-\omega_2).
\end{eqnarray}
According to noise correlations Eq.~(\ref{Noise correlation in momentum space}) and solution of
anti-Stokes mode Eq.~(\ref{Solutions of a and b in frequency domain}), we have
\begin{eqnarray}\label{Correlation relationships of anti-Stokes in momentum space}
	\langle \tilde{a}^{\dagger}(\omega_1) \tilde{a}(\omega_2) \rangle &=& \frac{ 2\pi g_{\rm om}^2 \Gamma_{\rm m} n_{\rm th} \delta(\omega_1-\omega_2) }
	{ \left | g_{\rm om}^2 + \left( \frac{\gamma_{\rm opt}}{2}+i(\omega_1+\Delta_{\rm as}) \right)
	 \left( \frac{\Gamma_{\rm m}}{2}+i(\omega_1+\Delta_{\rm ac}) \right)  \right |^2 }, \nonumber\\
	\langle \tilde{a}(\omega_1) \tilde{a}^{\dagger}(\omega_2) \rangle &=& 2\pi \frac{ \gamma_{\rm opt} \left( \frac{\Gamma_{\rm m}^2}{4} + (\omega_1+\Delta_{\rm ac})^2 \right)
	+ g_{\rm om}^2 \Gamma_{\rm m} (n_{\rm th}+1) }
	{ \left | g_{\rm om}^2 + \left( \frac{\gamma_{\rm opt}}{2}+i(\omega_1+\Delta_{\rm as}) \right)
		\left( \frac{\Gamma_{\rm m}}{2}+i(\omega_1+\Delta_{\rm ac}) \right)  \right |^2 } \delta(\omega_1-\omega_2), \nonumber\\
	\langle \tilde{a}^{\dagger}(\omega_1) \tilde{a}^{\dagger}(\omega_2) \rangle &=& 0, \nonumber \\
	\langle \tilde{a}(\omega_1) \tilde{a}(\omega_2) \rangle &=& 0.
\end{eqnarray}
Thus we have
\begin{eqnarray}\label{Displacement spectrum of anti-Stokes 2}
	&& \langle \tilde{X}_{a}^{\dagger}(\omega_1) \tilde{X}_a(\omega_2) \rangle \nonumber\\
	&=& \frac{ \langle \tilde{a}^{\dagger}(\omega_1) \tilde{a}(\omega_2) \rangle 
	+ \langle \tilde{a}(\omega_1) \tilde{a}^{\dagger}(\omega_2) \rangle
	+ \langle \tilde{a}^{\dagger}(\omega_1) \tilde{a}^{\dagger}(\omega_2) \rangle 
	+ \langle \tilde{a}(\omega_1) \tilde{a}(\omega_2) \rangle }{2} \nonumber\\
	&=& 2\pi \frac{ \frac{\gamma_{\rm opt}}{2}\left( \frac{\Gamma_{\rm m}}{2} + (\omega_1+\Delta_{\rm ac})^2 \right)
		+ \frac{\Gamma_{\rm m}}{2} g_{\rm om}^2 (2 n_{\rm th} + 1) }
		{ \left( g_{\rm om}^2 + \frac{\gamma_{\rm opt}\Gamma_{\rm m}}{4} - (\omega_1+\Delta_{\rm as})(\omega_1+\Delta_{\rm ac}) \right)^2 
			+ \frac{1}{4} \left( \gamma_{\rm opt}(\omega_1+\Delta_{\rm ac}) + \Gamma_{\rm m}(\omega_1+\Delta_{\rm as}) \right)^2 } \delta (\omega_1-\omega_2), \nonumber\\
\end{eqnarray}
By comparing Eq.~(\ref{Displacement spectrum of anti-Stokes 1}) and Eq.~(\ref{Displacement spectrum of anti-Stokes 2}),
the displacement spectrum of the $k$-th anti-Stokes mode can be given by
\begin{eqnarray}
	S_{XX}^{\rm as}(k,\omega) = \frac{ \frac{\gamma_{\rm opt}}{2}\left( \frac{\Gamma_{\rm m}}{2} + (\omega+\Delta_{\rm ac})^2 \right)
		+ \frac{\Gamma_{\rm m}}{2} g_{\rm om}^2 (2 n_{\rm th} + 1) }
	{ \left( g_{\rm om}^2 + \frac{\gamma_{\rm opt}\Gamma_{\rm m}}{4} - (\omega+\Delta_{\rm as})(\omega+\Delta_{\rm ac}) \right)^2 
		+ \frac{1}{4} \left( \gamma_{\rm opt}(\omega+\Delta_{\rm ac}) + \Gamma_{\rm m}(\omega+\Delta_{\rm as}) \right)^2 }.
\end{eqnarray}

\begin{figure}
		\centering
		\includegraphics[width=0.6\textwidth]{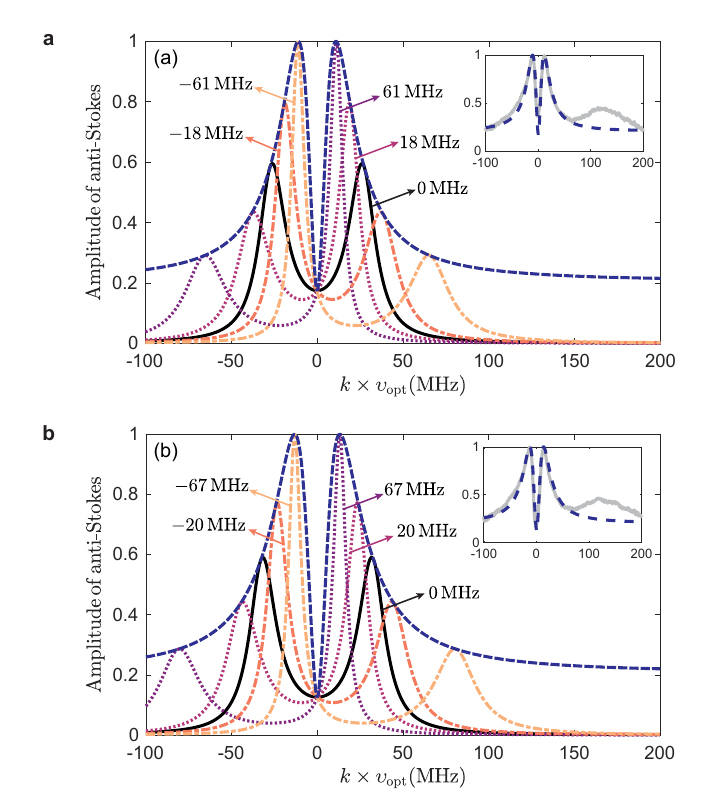} 
		\caption{{\textbf{Simulation results of spectra for the anti-Stokes wave.} (a)~The spectra of anti-Stokes wave with pump power of $\rm 0.83~W$ for various wave number $k$, 
			where dotted (dash-dot) lines denote the spectrum of $k$-th anti-Stokes mode with positive (negative) wave number $k$
			and the black solid line correspond to the phase-matching case ($k=0$). The dark blue dashed line shows the envelope of spectra
			for anti-Stokes modes with continuous wave number $k$. The inset indicates that experiment results (grey solid line) 
			agree well with theoretical results (dark blue dashed line). (b)~corresponds to the spectra of anti-Stokes wave with pump power of $\rm 1.21~W$.} }
		\label{SupMat_Theory} 
	\end{figure}

We show simulation results of spectra for anti-Stokes modes under the strong coupling regime in Fig.~\ref{SupMat_Theory} (a) and (b).
The dotted and dash-dot curves denote spectra of optical anti-Stokes modes for positive and negative wave number $k$, respectively.
The black solid curve corresponds to the phase-matching case, i.e., $k=0$. We can see that in the strong coupling regime, each anti-Stokes
mode exhibits the normal-mode splitting. When the wave number $k$ changes from negative (dash-dot curves) to positive values (dotted curves), 
there is an avoided crossing, which is demonstrated in Fig.~3 in the main text. The insets in Fig.~\ref{SupMat_Theory} (a) and (b) present
the simulation (blue dashed curves) and experiment (red solid curves) results for different pump powers, which are in agreement with each other.

Similarly, the displacement spectrum for the $k$-th acoustic mode can be expressed as
\begin{eqnarray}
	S_{XX}^{\rm ac}(k,\omega) = \frac{ \frac{\gamma_{\rm opt}}{2}\left( \frac{\Gamma_{\rm m}}{2} + (\omega+\Delta_{\rm ac})^2 \right)
		+ \frac{\Gamma_{\rm m}}{2} g_{\rm om}^2 (2 n_{\rm th} + 1) }
	{ \left( g_{\rm om}^2 + \frac{\gamma_{\rm opt}\Gamma_{\rm m}}{4} - (\omega+\Delta_{\rm as})(\omega+\Delta_{\rm ac}) \right)^2 
		+ \frac{1}{4} \left( \gamma_{\rm opt}(\omega+\Delta_{\rm ac}) + \Gamma_{\rm m}(\omega+\Delta_{\rm as}) \right)^2 }.
\end{eqnarray}
The eigenvalues of coupled anti-Stokes and acoustic modes with wave number $k$ can be given by
\begin{eqnarray}
	\omega_{\pm} = -\frac{\gamma_{\rm opt}+\Gamma_{\rm m}}{4} + i\frac{ \Delta_{\rm as} + \Delta_{\rm ac} }{2}
	 \pm i\sqrt{ g_{\rm om}^2 - \left( \frac{\gamma_{\rm opt}-\Gamma_{\rm m}}{4} - i \frac{ \Delta_{\rm as} - \Delta_{\rm ac} }{2} \right)^2 }.
\end{eqnarray}

\subsection*{Sample}

The sample used in the experiment is a 2.5~m highly nonlinear fiber (HNLF) placed inside a 4K Helium cryostat. The higher nonlinearity comes from the higher doping of Germanium in the core compared to single mode fiber (SMF). An HNLF has the same step-index, core-cladding structure as an SMF, allowing the sample to be spliced to standard APC connectors. The cladding and core diameters for both types of fiber is 125~µm and 8.2~µm \cite{HttpsWwwthorlabscomDrawings}, respectively. The loss of the sample in-coupling splice is -1.48~dB and of the out-coupling splice, -1.79~dB. In order to avoid heating because of the losses of the splices, these were placed outside of the cold plate of the cryostat, resulting in a cold section of the fiber estimated to be 1.5~m length. The parameters of the sample are given in Table~\ref{tab:sup_SampleBrillouinParameters}, comparing with standard SMF. The optical dissipation rate for this sample is $\gamma_{{\rm opt}}$~=~24~±~2~MHz, where $\gamma_{{\rm opt}}$ is defined as
\begin{equation}
    \gamma_{{\rm opt}}=\frac{\alpha_{\rm linear}c}{2\pi n},
\end{equation}
where $c$ is the speed of light in vacuum, $n$ the material refractive index and $\alpha_{\rm linear}$ the linear loss.

\begin{table*} 
	\centering
	\caption{\textbf{Brillouin parameters for different samples.}
		Comparison of Brillouin parameters between room temperature standard SMF and the HNLF sample used in this study at room and cryogenic temperatures. These parameters are obtained for $\lambda_{\rm pump}$~=~1550~nm. SMF parameters from \cite{kobyakovStimulatedBrillouinScattering2010a}.}
	\label{tab:sup_SampleBrillouinParameters}

	\begin{tabular}{c c c c c c c}
		\\
		\hline
		Fiber type~ & Temperature (K)~ & Length (cm)~ & $\Omega_{\rm B}/2\pi$ (GHz)~ & $\Gamma_{\rm m}$ (MHz)~ & $G_{\rm B}$ (W$^{-1}$m$^{-1}$)~ & $g_{\rm B}$ (W$^{-1}$m)~ \\
		\hline
		SMF  & 293 & -   & 10.87 & 20   & 0.14 & 1.19 · $10^{-11}$\\
		HNLF & 293 & 250 & 9.73  & 46.3 & 1.60 & 8.1 · $10^{-11}$\\
		HNLF & 4   & 150 & 9.60  & 9.91 & 6    & 3.2 · $10^{-10}$\\
		\hline
	\end{tabular}
\end{table*}

\subsection*{Thermal-noise-initiated measurement}

\begin{figure} 
	\centering
	\includegraphics[width=0.5\textwidth]{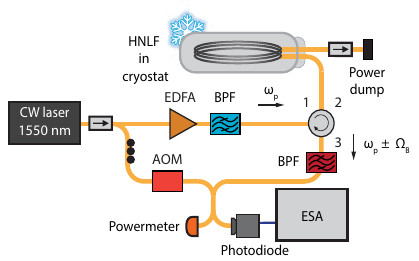}
	\caption{\textbf{Thermal-noise-initiated experiment.} Diagram of the experimental setup used to measure thermal-noise-initiated Brillouin-Mandelstam scattering via heterodyne detection. CW: continuous wave; AOM: acousto-optical modulator; EDFA: erbium-doped fiber amplifier; BPF: band-pass filter, ESA: electrical spectrum analyzer. The three black circles represent a fiber polarization controller.}
	\label{fig:SupMat_Setup_Spontaneous} 
\end{figure}

The setup used to perform a thermal-noise-initiated measurement via heterodyne detection is shown in Fig.~\ref{fig:SupMat_Setup_Spontaneous}.
The continuous emission of a narrow line laser with a wavelength $\lambda_{{\rm pump}}$~=~1550~nm is separated into two paths, pump and local oscillator (LO). The light in the pump arm is first amplified by a 5~W Erbium doped fiber amplifier (EDFA) and then filtered with a 0.8~nm tunable filter to remove the amplified spontaneous emission (ASE) coming from the EDFA. The pump is guided to the sample via a circulator, which redirects the back-reflected signal coming from the fiber into the detection part of the setup. As this signal will contain both the pump backreflection and the Brillouin resonance, a narrow-line tunable filter is used to selectively detect the Stokes or anti-Stokes signal. The leftover transmitted pump is dumped after the sample. An optical isolator is placed before the dump in order to reduce the back-reflected noise by 12 dB. Regarding the LO arm of the experiment, the frequency of the light is shifted 200~MHz by an acousto-optic modulator (AOM). This is allows to separate the two resonances in the detection device, which is particularly important for high pump power measurements in the cryostat. As the Stokes process enters the stimulated Brillouin-Mandelstam scattering (SBS) regime, the signal becomes so strong that it can not be fully filtered out. Frequency shifting the LO allows to measure the anti-Stokes signal without perturbations. The back-reflected signal and the LO are mixed in a 90:10 coupler and the interference directed to a high speed photodiode. A polarization controller in the LO arm is used to optimize the polarization overlap and therefore the signal. Finally, the transduced optical-into-electrical signal of the photodiode is sent and analyzed by an electrical spectrum analyzer (ESA).

\begin{figure} 
	\centering
	\includegraphics[width=0.8\textwidth]{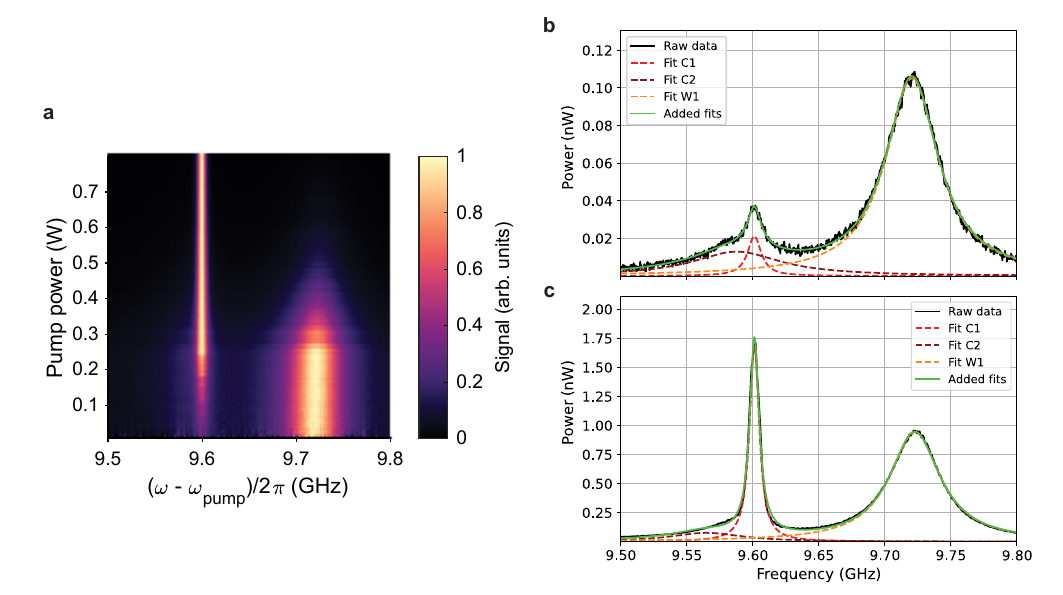} 
	\caption{\textbf{Thermal-noise-initiated measurement of Stokes resonances.} (a)~Normalized Stokes Brillouin response of the sample as a function of pump power. As the pump power is increased, the resonance coming from the 4~K section (peak at 9.5997~GHz) enters the stimulated regime and increases exponentially. The response of the room temperature part (peak at 9.7278~GHz) is only visible for low powers, comparatively. (b)~Spectrum of the full Stokes response of the sample at low input power. The signal centered at 9.6~GHz is composed of two peaks, a narrow 9.91~MHz one coming from the 4~K section of the fiber and a broad 77.2~MHz one coming from the transition to room temperature, at 190~K. (c)~Spectrum of the full Stokes response of the sample at higher input power. The spectrum is now dominated by the narrow peak coming from the 4~K section of the fiber.}
	\label{fig:SupMat_Spontaneous_BrillouinResponse_Stokes}
\end{figure}

\begin{figure} 
	\centering
	\includegraphics[width=0.8\textwidth]{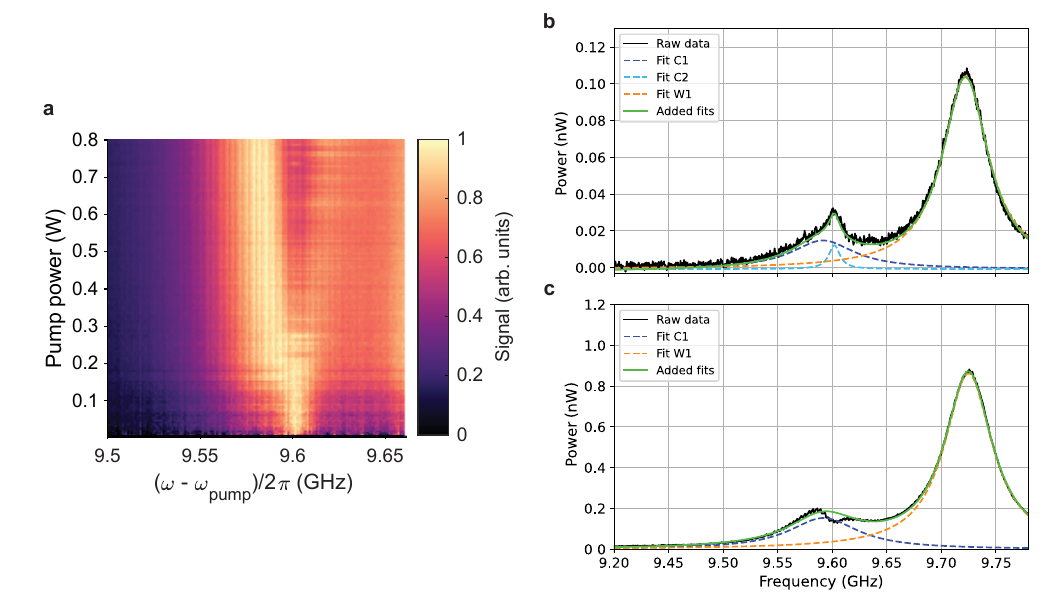}
	\caption{\textbf{Thermal-noise-initiated measurement of anti-Stokes resonances.} (a)~Normalized Anti-Stokes Brillouin response of the cryogenic sample as a function of pump power. As the pump power is increased, the resonance splits, corresponding to the transition into the strong coupling regime. (b)~Spectrum of the full anti-Stokes response of the sample at low input power. Same behavior as described in \ref{fig:SupMat_Spontaneous_BrillouinResponse_Stokes}~(b). (c)~Spectrum of the full anti-Stokes response of the sample at higher input power. As the pump power increases, the signal coming from the 4~K region of the sample dominates the 9.6~GHz spectrum, showing the strong coupling splitting.}
	\label{fig:SupMat_Spontaneous_BrillouinResponse_antiStokes}
\end{figure}

\begin{figure} 
	\centering
	\includegraphics[width=0.8\textwidth]{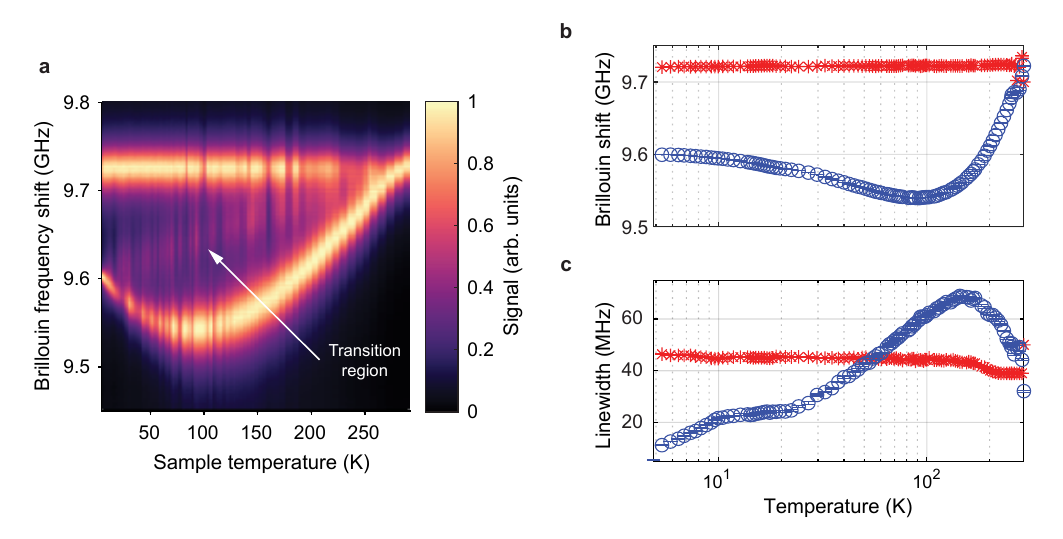} 
	\caption{\textbf{Stokes Brillouin response versus sample temperature.} (a)~Stokes response of the sample as a function of sample temperature. The resonance at the top, centered at ($\Omega_B/2\pi$~=~9.72~GHz), corresponds to the room temperature part of the fiber, and it does not shift as its temperature remains constant. On the lower part of the plot, the behavior of the sections of the fiber affected by the variation of temperature is shown. The most pronounced peak corresponds to the fiber section on the base plate of the cryostat. A faint third peak (indicated by the white arrow) is also present, coming from parts of the fiber in the transition. (b)~Brillouin shift of two most prominent peaks as a function of sample temperature. The room temperature part of the fiber (red stars) is unaffected by the change in temperature. The temperature-varying peak (blue circles) crosses the 9.6~GHz point twice, at T~=~191~K and at T~=~4~K. (c)~FWHM of the two most prominent peaks as a function of sample temperature. The room temperature part of the fiber (red stars) is unaffected by the change in temperature.}
	\label{fig:SupMat_Spontaneous_TempSweep} 
\end{figure}

The results for the experimental study of Brillouin-Mandelstam scattering in the sample at cryogenic temperatures are shown in Fig.~\ref{fig:SupMat_Spontaneous_BrillouinResponse_Stokes} and Fig.~\ref{fig:SupMat_Spontaneous_BrillouinResponse_antiStokes} . As explained in the previous section, not the whole HNLF is at low temperature, as sections of it are outside of the base plate in order to allow splicing for optical access. This results in three regions of the fiber: cold, room temperature and transition between the two. This is reflected in the spectra shown in Fig.~\ref{fig:SupMat_Spontaneous_BrillouinResponse_Stokes}~(a). Two main peaks are present, the cold one centered at $\Omega_{\rm B}/2\pi$~=~9.5997~±~0.0001~GHz and the room temperature one centered at $\Omega_{\rm B}/2\pi$~=~9.7278~±~0.0004~GHz. As the pump power is increased, the cold Stokes resonance will enter the stimulated Brillouin-Mandelstam scattering regime (SBS), increasing exponentially and narrowing from 9.91~±~0.07~MHz to 5.14~±~0.04~MHz. On the contrary, the anti-Stokes resonance broadens until it splits in two, meaning the strong coupling regime has been reached (Fig.~\ref{fig:SupMat_Spontaneous_BrillouinResponse_antiStokes}~(a)). Having a closer look at the cold part of the spectrum for low input power (Fig.~\ref{fig:SupMat_Spontaneous_BrillouinResponse_Stokes}~(b) and Fig.~\ref{fig:SupMat_Spontaneous_BrillouinResponse_antiStokes}~(b)), two peaks of different linewidths centered around 9.5997~GHz are present. A narrow 9.91~MHz peak, corresponding to the cryogenic part of the sample, and a broader 77.2~±~0.7~MHz peak. The source of this signal is the transition regions of the fiber, particularly of temperature around 190~K (Fig.~\ref{fig:SupMat_Spontaneous_TempSweep} (a)). Around that temperature, the frequency shift is the same as at 4~K (Fig.~\ref{fig:SupMat_Spontaneous_TempSweep} (b)), but the linewidth is higher (Fig.~\ref{fig:SupMat_Spontaneous_TempSweep} (c)) \cite{cryer-jenkinsBrillouinMandelstamScattering2025b}. 

The fact that these three regions of temperature are present in the sample is shown in Fig.~\ref{fig:SupMat_Spontaneous_TempSweep} (a), where three peaks are present in the spectrum, the transition region one being the faintest because of the short length of fiber that generates it. The reason why there is a length of fiber at this specific temperature long enough to generate a distinct signal could be the following. Two parts of the fiber of approximately 1~cm each are adhered with silicone tape to the radiation shield of the cryostat, to which they are thermalized. As the rest of the fiber is suspended in vacuum, these are the only parts outside of a smooth temperature gradient. Given the short length of these sections, their response is only non-negligible at very low pump powers, in which the totality of the Brillouin response of the fiber comes from spontaneous scattering. As they are at a higher temperature than the cold part, more phonons are present there, allowing for a scattered signal to be measured. In order to calculate $\Gamma_{\rm m}$ correctly they need to be considered in the analysis. Nonetheless, for higher pump power, the response of the fiber at 9.5997~GHz mainly originates from the cold part of the fiber (Fig.~\ref{fig:SupMat_Spontaneous_BrillouinResponse_antiStokes}~(c) and Fig.~\ref{fig:SupMat_Spontaneous_BrillouinResponse_Stokes}~(c)).

\subsection*{Pump-probe measurement}

\begin{figure} 
	\centering
	\includegraphics[width=0.5\textwidth]{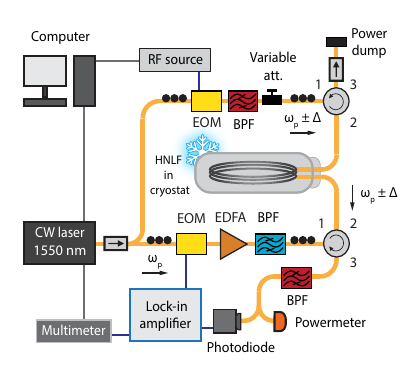}
	\caption{\textbf{Pump-probe experiment.} Diagram of the setup used to measure Brillouin-Mandelstam scattering via a pump-probe experiment with lock-in amplifier detection. EOM: electro-optical modulator; Variable att.: variable attenuator; RF: radio frequency. The remaining abbreviations same as Fig.~\ref{fig:SupMat_Setup_Spontaneous}.}
	\label{fig:SupMat_Setup_Seeded} 
\end{figure}

The setup used to perform the pump-probe experiment is shown in Fig.~\ref{fig:SupMat_Setup_Seeded}. The output of the laser described in the previous section is divided into two paths, pump and probe. The light of the pump arm is modulated by an electro-optical modulator (EOM) before amplification. The modulation frequency is given by a lock-in amplifier. The light of the probe arm is modulated by an EOM connected to an RF source, generating thus two sidebands $\omega_{{\rm probe}}~=~\omega_{{\rm pump}}\pm\Omega_{\rm B}+\Delta_{{\rm probe}}$. These sidebands are swept around the resonance frequency $\Omega_{\rm B}$. A narrowline bandpass filter is placed after the EOM to select one sideband and probe only one of the Brillouin processes. The light of each arm is coupled into the sample in opposite directions via circulators. To maximize the interaction between the different waves, polarization controllers are placed on each arm. The port 3 of the probe circulator is connected to an insulator followed by a power dump to get rid of the remaining pump light. The signal coming out of the port 3 of the pump circulator contains the probe frequency after the Brillouin interaction in the fiber plus the spontaneous scattering generated by the pump. To decrease the noise in the measurement, the back-reflection is filtered by a narrow bandpass filter. This signal is directed to a photodiode connected to the lock-in amplifier, where the resulting electrical signal is demodulated to obtain the probe power as a function of time. The amplified electrical output of the lock-in is measured by a multimeter. Both the RF-source and the multimeter are controlled by a computer, allowing to measure the resulting Brillouin spectrum after the sweep without the need of a frequency reference. It was not observed that the low power probe disturbs the strongly coupled system.

\subsection*{Temperature dependence of strong coupling threshold}

\begin{figure}
	\centering
	\includegraphics[width=0.8\textwidth]{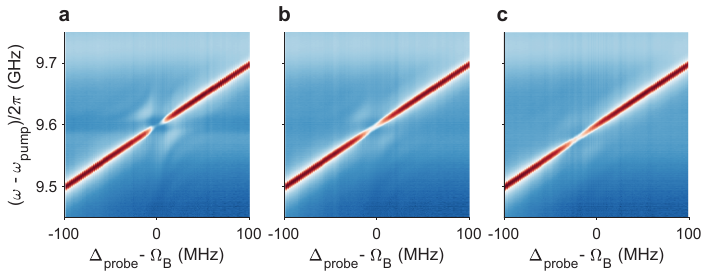}
	\caption{\textbf{Temperature dependence of strong coupling regime.} (a)~Forced detuning measurement at $T$~=~4~K, $P_{{\rm pump}}$~=~1.35~±~0.1~W ($g_{{\rm om}}$~=~70~±~2~MHz). As the system is in the strong coupling regime, it shows a clear avoided crossing. (b)~Forced detuning measurement for $T$~=~10~K, same pump power as (a). The avoided crossing is not clearly visible, as the system enters the weak coupling regime. (c)~Same as (b) with $T$~=~19~K.}
	\label{fig:SupMat_ForcedDetuning_vs_T} 
\end{figure}

As it was stated when describing the sample, we are able to reach the strong coupling regime thanks to the balance of losses, Brillouin gain and power handling in our platform. Working at cryogenic temperature is beneficial, as the acoustic dissipation rate ($\Gamma_{\rm m}$) decreases significantly and the Brillouin gain $G_{\rm B}$, on which the effective coupling strength depends, increases comparing to room temperature. Working at $T$~=~4~K, the lowest temperature accessible in our experiment, proved to be our optimal point to perform strong coupling measurements (Fig.~\ref{fig:SupMat_ForcedDetuning_vs_T}~(a)). A temperature increase moves the system into the weak coupling regime. This can be seen in Fig.~\ref{fig:SupMat_ForcedDetuning_vs_T}~(b) and (c), where the pump power is the same as in (a), but the temperature is $T$~=~10~K and $T$~=~19~K, respectively. In these two cases, the avoided crossing is less significant than at 4~K. The ideal working will differ in a different material and type of device.

\end{document}